\documentclass[11pt,a4paper]{article}
\usepackage{jheppub}
\usepackage[utf8]{inputenc}
\usepackage{graphicx}
\usepackage{epsfig}
\usepackage{subfigure}
\usepackage{color}
\usepackage{comment}
\usepackage{amsmath}
\usepackage{mathtools}
\usepackage{tabularx}
\usepackage{tabu}
\usepackage{hhline}
\usepackage{multirow}
\usepackage{bm}

\newcommand{\be}{\begin{equation}}
\newcommand{\ee}{\end{equation}}

\newcommand{\dd}{\mathrm{d}}
\providecommand{\f}[2]{\frac{{#1}}{{#2}}}
\title{Vacuum decay constraints on the Higgs curvature coupling from inflation}

\newcommand{\Lagr}{\mathcal{L}}

\def\Nbub{\langle {\cal N}\rangle}

\author[a]{Andreas Mantziris} 
\author[b,c]{\!, Tommi Markkanen}
\author[a]{and Arttu Rajantie}

\affiliation[a]{Department of Physics, Imperial College London, London, SW7 2AZ, United Kingdom}
\affiliation[b]{Laboratory of High Energy and Computational Physics, National Institute of Chemical Physics and Biophysics, R\"avala pst. 10, Tallinn, 10143, Estonia}
\affiliation[c]{Helsinki Institute of Physics, P.O. Box 64, FIN-00014 University of Helsinki, Finland}

\emailAdd{a.mantziris18@imperial.ac.uk}
\emailAdd{tommi.markkanen@kbfi.ee}
\emailAdd{a.rajantie@imperial.ac.uk}

\abstract{We derive lower bounds for the Higgs-curvature coupling from vacuum stability during inflation in three inflationary models: quadratic and quartic chaotic inflation, and Starobinsky-like power-law inflation. In contrast to most previous studies we take the time-dependence of the Hubble rate into account both in the geometry of our past light-cone and in the Higgs effective potential, which is approximated with three-loop renormalisation group improvement supplemented with one-loop curvature corrections. We find that in all three models, the lower bound is $\xi\gtrsim 0.051\ldots 0.066$ depending on the top quark mass. We also demonstrate that vacuum decay is most likely to happen a few $e$-foldings before the end of inflation.}

\begin{document}
\maketitle
\section{Introduction} \label{sec:Introduction}

The experimentally measured mass \cite{Chatrchyan:2012ufa, Aad:2012tfa} of the Higgs boson lies in a range within which the Higgs self-interaction does not diverge below the Planck scale \cite{Degrassi:2012ry,Buttazzo:2013uya,Bednyakov:2015sca}. This has attracted significant interest in the past \cite{Hung:1979dn,Sher:1993mf,Casas:1996aq,Isidori:2001bm,Ellis:2009tp,EliasMiro:2011aa,Lebedev:2012zw,Branchina:2013jra} and implies that the Standard Model (SM) of particle physics might be sufficient to describe our universe up {to Planck scale energies where} quantum gravity effects become significant. {This means that the SM can be used as a consistent minimal model for describing the early Universe and addressing open cosmological questions.} 

The current parameters of the SM, in particular the masses of the Higgs boson and the top quark, suggest that the Universe lies currently in a metastable electroweak vacuum, however with an extremely small decay rate that makes a collapse into the true minimum very unlikely \cite{Salvio:2016mvj,Rajantie:2016hkj,Chigusa:2017dux,Chigusa:2018uuj,Chigusa:2020jbn,Espinosa:2020qtq}. This means that in principle, given a long enough time interval, the Higgs field will eventually decay to its true vacuum state
through the nucleation of a bubble of true vacuum. Inside the bubble, space-time is described by quantum gravity \cite{Espinosa:2015qea}, but for all practical purposes in the framework of the SM, it is considered to collapse into a singularity. The bubble will then grow at the speed of light, destroying everything in its way \cite{Callan:1977pt,Coleman:1980aw}. From the observation that the Universe around us is still in the metastable state, we can therefore conclude that no such bubble nucleation event took place inside our past light-cone.
In the early Universe, however, the probability of such an event could have been close to unity, which allows us to constrain fundamental theories and their parameters leading to a wide variety of physical implications reviewed in Ref.~\cite{Markkanen:2018pdo}, where we refer the reader for more details and references.

There is already a substantial body of literature investigating implications from vacuum stability during inflation \cite{Espinosa:2018mfn,Lebedev:2012sy,Kobakhidze:2013tn,Fairbairn:2014zia,Hook:2014uia,Kamada:2014ufa,Espinosa:2015qea,Kearney:2015vba,East:2016anr,Enqvist:2014bua,Bhattacharya:2014gva,Herranen:2014cua,Czerwinska:2015xwa,Rajantie:2016hkj,Czerwinska:2016fky,Rajantie:2017ajw,Markkanen:2018bfx,Rodriguez-Roman:2018swn,Rusak:2018kel,Fumagalli:2019ohr,Jain:2019wxo,Hertzberg:2019prp,Lalak:2020dao,Adshead:2020ijf} and reheating \cite{Herranen:2015ima,Kohri:2016wof,Gross:2015bea,Ema:2016kpf,Enqvist:2016mqj,Ema:2017loe,Postma:2017hbk,Figueroa:2017slm,Croon:2019dfw} and possible cosmological signatures from non-fatal scenarios as well as effects from black holes \cite{Kawasaki:2016ijp,Burda:2015isa,Burda:2016mou,Espinosa:2017sgp,Kohri:2017ybt,Espinosa:2018euj,Franciolini:2018ebs,Cline:2018ebc,Espinosa:2018eve,Hook:2019zxa,Hayashi:2020ocn}. In particular, we highlight the recent analysis of inflationary vacuum stability in Ref.~\cite{Fumagalli:2019ohr}, which addressed the time-dependence of the background as well as the influence from  Planck-suppressed derivative operators.

The bubble nucleation probability depends sensitively on the non-minimal Higgs-curvature coupling $\xi$, which is a renormalisable parameter of the SM in curved space-time \cite{Chernikov:1968zm,Callan:1970ze,Bounakis:2017fkv,Markkanen:2018bfx}. However, its value is very difficult to measure experimentally in the present-day Universe because of the low space-time curvature. Therefore, the constraints from cosmological vacuum instability are many orders of magnitude stronger than those from other measurements \cite{Markkanen:2018pdo}.

Most of the existing literature has, nonetheless, approximated the space-time during inflation with a de Sitter (dS) space-time. That means that the Hubble rate is treated as a constant free parameter. In this work, we consider the question in the context of actual inflationary models in which the Hubble rate is time dependent and, once a model is chosen, the parameters are determined by cosmic microwave background (CMB) observations. The choice of the inflationary model affects the bubble nucleation probability in two ways: Because of the time-dependence of the Hubble rate, the space-time curvature is different, which affects the nucleation rate per unit space-time volume.{ We compute this using the renormalisation group improved effective Higgs potential with three-loop running~\cite{Chetyrkin:2012rz,Bezrukov:2012sa}, pole matching as described in Ref.~\cite{Bezrukov:2009db} and where crucially the effective potential is calculated on a curved background to one-loop order as given in Ref.~\cite{Markkanen:2018bfx}.} The model choice also determines the geometry of the past light-cone, which is different from dS. We incorporate both of these effects and find the constraints on the Higgs-curvature coupling $\xi$ arising from vacuum stability during inflation in three specific inflationary models, considering also the dependence on the top quark mass.

The format of this paper is as follows. In section \ref{sec:Effective potential for the SM Higgs}, we focus on the Higgs effective potential, going from the simplest tree level case in flat space to the current state-of-the art with 3-loops and one-loop curvature corrections. In section \ref{sec:Bubble nucleation during inflation}, there is a  description of the context and the mathematical treatment of the electroweak vacuum instability along with a brief overview of the inflationary models considered. Afterwards in section \ref{sec:Results}, we present our findings for the constraints on the non-minimal coupling $\xi$ for the different inflationary models and a range of top quark masses. We also discuss the time of the nucleation event and its connection with the duration of inflation. Finally, in section \ref{sec:Conclusions} we delineate the context of the study, highlighting its differences from the past literature. We also provide an overview of our results, where we underline their cosmological implications.

\section{Effective potential for the SM Higgs} \label{sec:Effective potential for the SM Higgs}
\subsection{Vacuum instability in Minkowski space}
The SM is an effective quantum field theory (QFT) and as such the effective interactions between particles that we observe, are related to the energy regime at which they take place. This is treated formally in the context of renormalization. Via this procedure we see that our renormalized parameters are functions of the renormalization scale $\mu$, which corresponds to the energy scale of the physics one is probing. If a renormalized coupling diverges (i.e., has a Landau pole) at energies lower than the Planck scale, then this means that our theory is not valid at those energies and thus we need to think of ways to fix it, such as beyond the Standard Model (BSM) physics. \cite{Markkanen:2018pdo, Cottingham:2007zz, Stopyra:2018cjy, Lancaster:2014pza, Gies:2017ajd}

Specifically for our case, when the Higgs field interacts with bosons, its self-coupling $\lambda (\mu)$ is increased, whereas fermionic interactions decrease it. The size of these contributions scales with the mass of the corresponding particle. In the SM, the top quark and the Higgs are by far the heaviest of the fermions and bosons respectively and therefore dominate the contributions to the self-coupling than the rest of the particle spectrum. Hence, if either one is significantly more massive than the other, then $\lambda$ diverges to $\pm \infty$ accordingly. This implies that the only way for the SM to remain valid until the Planck scale would be for the two masses to be comparable with one another, in order for the two competing contributions to cancel out. As it turns out, their experimentally measured masses \cite{Zyla:2020zbs} suggest exactly that. This realization acts as a strong constraint on BSM theories that would disrupt the balance between the two masses by affecting $\lambda$'s dependency on $\mu$. \cite{Markkanen:2018pdo}

According to the calculation reviewed in Ref.~\cite{Markkanen:2018pdo}, the Higgs self-coupling turns negative above approximately $10^{10}$ GeV resulting in an additional vacuum state of lower energy. This makes the current vacuum state that the Higgs field resides in to be metastable, and thus prone to vacuum decay via quantum tunneling. This process induces the formation of a bubble of true vacuum expanding with velocity close to the speed of light, destroying the universe as we know it in its path. The true vacuum may or may not be bounded from below, but in the context of this study, it is not relevant whether it is or not, because we focus on the bubbles that would form during the tunneling process. This is the case because we are interested in the possible signatures evident in our false vacuum Universe and not on the specifics of the exotic physics inside the true vacuum bubble, which would obviously depend on the form of the potential. \cite{Markkanen:2018pdo, Cottingham:2007zz, Degrassi:2012ry, Buttazzo:2013uya, Rojas:2015yzm, Stopyra:2018cjy}

 The vacuum decay can involve both thermal and quantum fluctuations to surpass the barrier and we quantify the probability of decay via the decay rate $\Gamma$. This is a function of space-time and also evidently of $m_h$ and $m_t$. For an infinitely old universe, even the most infinitesimal decay rate would render it incompatible with ours. Today, our measurements indicate that we are in the metastable vacuum with a decay rate that requires more time than the age of the universe for the process to occur \cite{Espinosa:2018mfn}. This result has two important implications. Firstly, it acts as a reality check for SM extensions, which should abide by this long-lasting false vacuum. Secondly, it places constraints in our early universe theories, where a higher decay rate was favoured, as the metastable vacuum has managed to survive through its various epochs. \cite{Markkanen:2018pdo, Stopyra:2018cjy} 

On a flat background often a reasonable approximation for the renormalization group improved effective potential is~\cite{Degrassi:2012ry}
\be{V_{\rm H}(h)\approx\f{\lambda(h)}{4}h^4\,,}\label{eq:RGIf}
\ee
with the choice $\mu=h$ as the renormalization group (RG) scale for the running four-point coupling. It should be solved to as high a precision as is practically feasible in order to accurately capture the running. The current state-of-the-art calculation~\cite{Bednyakov:2015sca} making use of two-loop matching conditions, three-loop RG evolution and pure QCD corrections to four-loop accuracy leads to an instability around the scale $\mu_{\Lambda}=1.60 \times 10^{10}$ GeV for the central values of the top quark and Higgs masses. It is however important to note that due to the experimental and theoretical inaccuracies, in particular in defining the top quark mass, absolute stability of the vacuum is still a viable possibility.

Direct loop corrections to the effective potential are neglected in Eq.~(\ref{eq:RGIf}), which may therefore be a poor approximation in some cases~\cite{Markkanen:2018pdo}. This can be remedied either by including the quantum corrections in Eq.~(\ref{eq:RGIf}) or by choosing explicitly the RG scale, instead of $\mu=h$, such that the quantum correction vanishes.

\subsection{Tree-level curvature corrections} 

During inflation, space-time was highly curved, and therefore the Minkowski calculation of the vacuum decay rate is not applicable. A detailed calculation of the decay rate in a general curved space-time would be very difficult, and therefore we approximate it locally with a dS space. The tunnelling process from false to true vacuum can be solved classically yielding solutions called instantons. The vacuum decay rate is then determined by the action of the Coleman-de Luccia instanton~\cite{Coleman:1980aw}. At sufficiently high Hubble rates, it approaches the much simpler Hawking-Moss instanton~\cite{Hawking:1981fz,Rajantie:2017ajw}, whose action difference is
\begin{eqnarray}
    B_{\rm HM}(R) \approx \frac{384 \pi^2 \Delta V_{\rm H} }{R^2}  \, ,
    \label{eq:BHM}
\end{eqnarray} 
where $\Delta V_{\rm H} = V_{\rm H}(h_{\rm bar}) - V_{\rm H}(h_{\rm fv})$ is the height of the potential barrier~\cite{Markkanen:2018pdo, Hawking:1981fz, Linde:1998gs}. This is the approximation we will use throughout this paper. It results in a reasonably good first approximation for the form of the decay rate as
\begin{eqnarray}
    \Gamma_{\rm HM}(R) &\approx & \left(\f{R}{12}\right)^2 e^{-B_{\rm HM}(R)}  \, ,
    \label{eq:Gamma}
\end{eqnarray}
where the prefactor is justified by dimensional arguments. For light fields, $m\ll H$, this agrees with the stochastic formalism~\cite{Espinosa:2007qp}.

Non-zero space-time curvature also affects the effective potential of the Higgs field. At tree level, it enters through the Higgs-curvature coupling $\xi$.\footnote{We use the sign convention in which the conformal value is $\xi=1/6$ and $(-, -, -)$ for the metric and curvature tensors according to \cite{Misner:1974qy}.} In this approximation, the Lagrangian for the Higgs field in curved space-time (in the Jordan frame) is given by \begin{eqnarray}
 \Lagr =  \frac{M_P^2}{2}R + \frac{1}{2}g^{\mu \nu} (\partial_{\mu} h) (\partial_{\nu} h)  - V_{\rm H} (h, R) \, ,
\end{eqnarray}
where the first term corresponds to the standard Einstein-Hilbert term with $M_P = (8\pi G)^{-1/2} \approx 2.435 \times 10^{18}$ GeV being the reduced Planck mass, the second is the usual kinetic term and the third is the effective curvature-dependent Higgs potential given at tree level by
\begin{eqnarray}
 V_{\rm H}(h,R) = \frac{\xi}{2}R h ^2 + V_{\rm H} (h),
\end{eqnarray}
where the first term couples the Higgs field with curvature and acts as a mass term~\cite{Markkanen:2017dlc}, and $V_{\rm H}(h)$ corresponds to a general flat space-time potential from QFT. As we will see, the relevant values of $\xi$ for our analysis are low, $\xi\ll 1/6$, and therefore the Higgs field remains light and we should be able to trust Eq.~(\ref{eq:Gamma}).

To understand this effect, let us consider constant $\xi$ and $\lambda<0$, which is a reasonable approximation for the Higgs potential at field values $h\gg 10^{10}~{\rm GeV}$. We may then write
\begin{eqnarray}
    V_{\rm H}(h, R) = \frac{\xi}{2} R h^2 - \frac{|\lambda|}{4}h^4  \,  ,
\end{eqnarray}
where the value of the potential at the top of the barrier is then
\begin{eqnarray}
    V_{\rm H}(h_{\mathrm{bar}}, R) =  \frac{\xi^2 R^2}{4|\lambda|} \,,
\end{eqnarray} 
resulting in the action difference via Eq.~(\ref{eq:BHM}),
\begin{eqnarray}
    B_{\rm HM} \approx \frac{96 \pi^2 \xi^2}{|\lambda|} \, .
    \label{eq:BHM-approx}
\end{eqnarray}
We can see that in this approximation the action is actually independent of space-time curvature, and that it is an increasing function of the curvature coupling $\xi$. This suggests that a sufficiently high value of $\xi$ will prevent vacuum decay during inflation and that, conversely, vacuum stability provides a lower bound on its value.

\subsection{One-loop curvature corrections}
Beyond tree level, space-time curvature also enters the effective potential through loop corrections. The effective potential for the full SM on a curved background to 1-loop order was calculated in Ref.~\cite{Markkanen:2018bfx}. In dS space it has the expression
\begin{eqnarray}
V_{\rm H}(h, \mu, R)=-\f{m^2}{2}h^2+\f{\xi}{2}Rh^2+\f{\lambda}{4}h^4+V_\Lambda-\kappa R+\frac{\alpha}{144} R^2 + \Delta V_{\rm loops} \,,
\label{eq:VeffSMdS}
 \end{eqnarray}
where the loop contribution can be parametrized as
\begin{eqnarray}
     \Delta V_{\rm loops} (h, \mu, R) = \frac{1}{64\pi^2} \sum\limits_{i=1}^{31}\bigg\{ n_i\mathcal{M}_i^4 \bigg[\log\left(\frac{|\mathcal{M}_i^2 |}{\mu^2}\right) - d_i \bigg] +\frac{n'_i}{144}R^2\log\left(\frac{|\mathcal{M}_i^2 |}{\mu^2}\right)\bigg\}  \,, \,\,\,\,\,\,
\end{eqnarray}
having suppressed all implicit dependence on the renormalization scale $\mu$ and summing over all degrees of freedom of the SM. The various terms can be found in Section 5 of Ref.~\cite{Markkanen:2018bfx}, where we note that we are using $\zeta_i = 1$ in this study.\footnote{Regarding the value of the potential at the top of the barrier, we have checked that there is no significant difference between using gauge fixings $\zeta_i = 0$ and $\zeta_i = 1$ in the loop contribution.} It is necessary to highlight that we will be neglecting the mass term for the Higgs because it is negligible compared to the scales of the Hubble rates for the models we study. Setting $m = 0$ implies that we can also disregard all dimensionful couplings of our theory, having a renormalisation group flow fixed point at $m = 0, V_{\Lambda} = 0, \kappa = 0$ \cite{Hardwick:2019uex}. Hence, we end up with 
\begin{eqnarray}
V_{\rm H}(h,\mu, R)=\f{\xi(\mu)}{2}Rh^2+\f{\lambda(\mu)}{4}h^4+\frac{\alpha(\mu)}{144} R^2 + \Delta V_{\rm loops} (h, \mu, R)  \, .
\label{eq:Higgs-Potential}
\end{eqnarray}

We wish to eliminate the dependence on the scale $\mu$ in Eq.~(\ref{eq:Higgs-Potential}) via Renormalization Group Improvement (RGI), where we fix $\mu=\mu_*(h,R)$ choosing $\mu_*$ in such a way as to result in null loop corrections to the potential~\cite{Ford:1992mv},
i.e. as a solution of
\begin{equation}
\Delta V_{\rm loops} (h, \mu_*, R) = 0.
\label{eq:dcond}
\end{equation}
This leads to the RGI effective potential
\begin{eqnarray}
    V_{\rm H}^{\rm RGI}(h, R) = \frac{\xi(\mu_*(h, R))}{2}  R h ^2 +\frac{\lambda(\mu_*(h, R))}{4}  h^4 + \frac{\alpha(\mu_*(h,R))}{144} R^2 \, ,
    \label{eq:RGIPot}
\end{eqnarray}
where because of the scale choice, there is no direct loop contribution, which implies a well-defined loop expansion\footnote{In Eq.~(\ref{eq:RGIPot}) $h$ refers to the field renormalized at a scale $\mu_*$, which is related to the field $h_0$ renormalized at some fixed physical scale $\mu_0$ via the anomalous dimension $\gamma$ as
\begin{equation}
h=h_0\exp\bigg(-\int_{0}^{\log\left(\frac{\mu_*}{\mu_0}\right)}\gamma(t)\dd t\bigg).
\end{equation}
Arguably, here the relevant physical quantity is the field renormalized at the electroweak scale instead of $h$. The Hawking-Moss instanton however is independent of this subtlety since it does not affect the barrier height, and for simplicity we may then perform our calculation in terms of $h$.}.

It is worth pointing out that a simpler scale choice of the form $\mu^2 = a h^2 + b R$, where $a$ and $b$ are constants, was suggested in \cite{Herranen:2014cua} and has since widely been used in the literature, see e.g.~Ref.~\cite{East:2016anr,Franciolini:2018ebs,Fumagalli:2019ohr,Cheong:2019vzl}. With that choice, the direct loop corrections do not cancel exactly and should therefore be included in the effective potential for full accuracy. On the other hand, it avoids the issue that in some cases Eq.~(\ref{eq:dcond}) does not have a continuous solution covering all field values~\cite{Markkanen:2018bfx}. In the current case, that issue does not arise and therefore we use the exact scale choice (\ref{eq:dcond}).

The beta functions for the SM non-gravitational couplings are well known~\cite{Bezrukov:2009db}, in some cases up to three loops and therefore we provide here explicitly only the 1-loop couplings associated with curvature, the non-minimal coupling and $\alpha$ as a reference \cite{Herranen:2014cua},
\begin{align}
16\pi^2 \frac{d \xi}{d\ln\mu}=
    16 \pi^2 \beta_{\xi} &= \left( \xi - \frac{1}{6} \right) \left( 12 \lambda + 6 y_t^2 - \frac{3}{2}g'^2 -\frac{9}{2} g^2 \right)  \, ,
    \label{eq:betaC1}\\
    16\pi^2 \frac{d \alpha}{d\ln\mu}=
  {16\pi^2}  \beta_{\alpha}&=288 \xi ^2-96 \xi -\frac{1751}{30} \label{eq:betaC2} \, .
\end{align}
Assuming given electroweak-scale values $\xi_{\rm EW}=\xi(\mu_{\rm EW})$ and $\alpha_{\rm EW}=\alpha(\mu_{\rm EW})$, these equations determine their scale-dependence, which enters Eq.~(\ref{eq:RGIPot}). The differential equations above are sensitive to the initial condition of $\xi$ but not of $\alpha$ since the right-hand sides on Eq.~(\ref{eq:betaC1}) and Eq.~(\ref{eq:betaC2}) are independent of $\alpha$. Although we will not be limiting our analysis to strict dS space, it is still a good approximation to make use of the dS form of (\ref{eq:VeffSMdS}): the higher order curvature invariants $R^2$, $R_{\mu\nu}R^{\mu\nu}$ and $R_{\mu\nu\alpha\beta}R^{\mu\nu\alpha\beta}$ couple to the Higgs only via loop corrections \cite{Markkanen:2018bfx} and reduce to a single term at the dS limit, indicating that our approximation captures the leading contribution, which we have checked is already a small contribution.

Solving Eq. (\ref{eq:dcond}) and calculating the barrier height of (\ref{eq:RGIPot}) require us to incorporate the entire SM particle spectrum. Firstly, we need to obtain the running of $\lambda, y_t, g', g$ according to the corresponding beta functions and the accompanying pole-matching. We perform this numerical task by using the publicly available Mathematica code\footnote{\label{foot:code} By Fedor Bezrukov, available at \href{http://www.inr.ac.ru/~fedor/SM/}{\textcolor{blue}{{http://www.inr.ac.ru/$\sim$\! fedor/SM/}}}.}, which is based on Refs.~\cite{Bezrukov:2009db, Bezrukov:2012sa, Chetyrkin:2012rz}. The code takes as input parameters the fine structure constant $\alpha_S$ and the masses of the Higgs boson $m_h$ and the top quark $m_t$ renormalised at the electroweak scale, and calculates the running of the SM parameters. After this, we obtain $\xi(\mu)$ and $\alpha(\mu)$ by solving the remaining beta functions (\ref{eq:betaC1}) and (\ref{eq:betaC2}). Finally, we can calculate the maximum of Eq.~(\ref{eq:RGIPot}) after having obtained $\mu_*$ via Eq.(\ref{eq:dcond}), including the masses of all the SM particles. The overview of all the input values along with the evaluated SM couplings is shown in Table \ref{tab:param_table}. 

\begin{table}[t]
\begin{center}
	\scalebox{0.95}{
	\begin{tabular}{|c|c|c|c|}
	\hline
	    & \multicolumn{3}{|c|}{Masses [GeV]}     \\ \hline
	 Higgs & \multicolumn{2}{|c|}{$m_h = 125.10$} & $v=246.22$ \\ \hline
	\multirow{2}{4em}{Quarks} &  $m_t = 172.76$ & $m_s = 93 \times 10^{-3}$& $m_u = 2.16 \times 10^{-3}$   \\  \cline{2-4}
	 & $m_b = 4.18$ & 	$m_c = 1.27$ & 	$m_d = 4.67 \times 10^{-3}$  \\  \hline
	Leptons & $m_{\tau} = 1.77686$ & $m_{\mu} = 105.6583745 \times 10^{-3}$ & $m_e = 510.9989461 \times 10^{-6}$ \\ \hline
	& \multicolumn{3}{|c|}{Dimensionless couplings} \\ \hline
	gauge couplings & $\alpha_S = 0.1179$ & $\bm{g=0.648382}$& $\bm{g'=0.358729}$   \\ \hline
		couplings  &  & $\bm{\lambda = 0.126249}$ & $\bm{y_t =0.934843}$  \\ \hline
	\end{tabular}}
\end{center}
	\caption{{Experimental values of the SM particle masses and couplings \cite{Zyla:2020zbs} at $\mu_{\rm EW}$ used for the RGI of the Higgs potential according to Eqs.~(\ref{eq:dcond})-(\ref{eq:RGIPot}). The couplings in bold are calculated by the SM code\textsuperscript{\ref{foot:code}} subject to the input of $m_h,m_t$ and $a_S$.}}
	\label{tab:param_table}
\end{table}

\begin{figure}[t]
    \centering
    \includegraphics[scale=0.67]{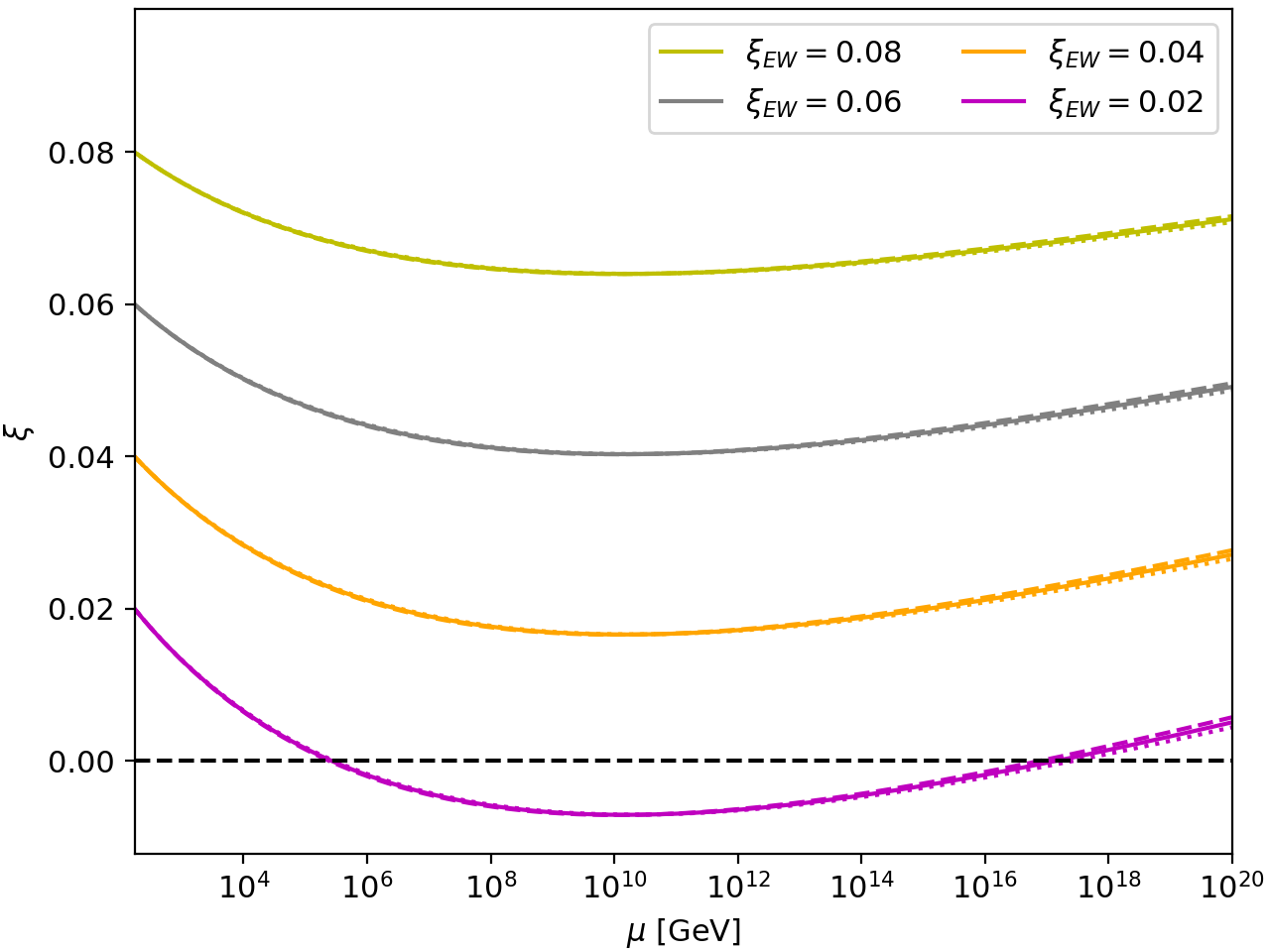}
    \caption{The running of non-minimal coupling $\xi(\mu)$ with various boundary conditions $\xi_{\rm EW}$ for top quark mass $m_t = (172.76 \pm 0.6)$ GeV. The solid lines correspond to the central value of $m_t$ for each case, while the dashed and the dotted ones correspond to $ m_t \pm 2 \sigma$, respectively.}
    \label{fig:ksis-mu}
\end{figure}

 Fig.\ref{fig:ksis-mu} shows the running of $\xi(\mu)$ according to Eq.~(\ref{eq:betaC1}) for a range of boundary conditions at $\mu_{\rm EW}$, where we can see that below $\xi_{\rm EW} \approx 0.03$, $\xi$ switches sign as it runs resulting in a negative term in the potential (\ref{eq:RGIPot}). This can potentially destabilize the Higgs vacuum depending on how the other couplings in Eq.~(\ref{eq:RGIPot}) run. Therefore, for all practical purposes in this study, we limit ourselves to the values of $\xi_{\rm EW}$ for which $\xi(\mu)$ remains positive and the Higgs potential resides ``safely'' in the meta-stability region. It is also evident that even a $2 \sigma$ deviation in the top quark mass does not affect the running of $\xi$ significantly at the energy scales of interest. In particular, it seems to produce a more observable effect beyond scales of order $10^{18}$ GeV and as $\xi_{\rm EW}$ gets smaller. We have not considered any deviation in the Higgs' mass and we kept it fixed at its central value throughout our calculations, because of its smaller experimental uncertainty compared to the top quark's.

\begin{figure}[b]
    \centering
    \includegraphics[scale=0.68]{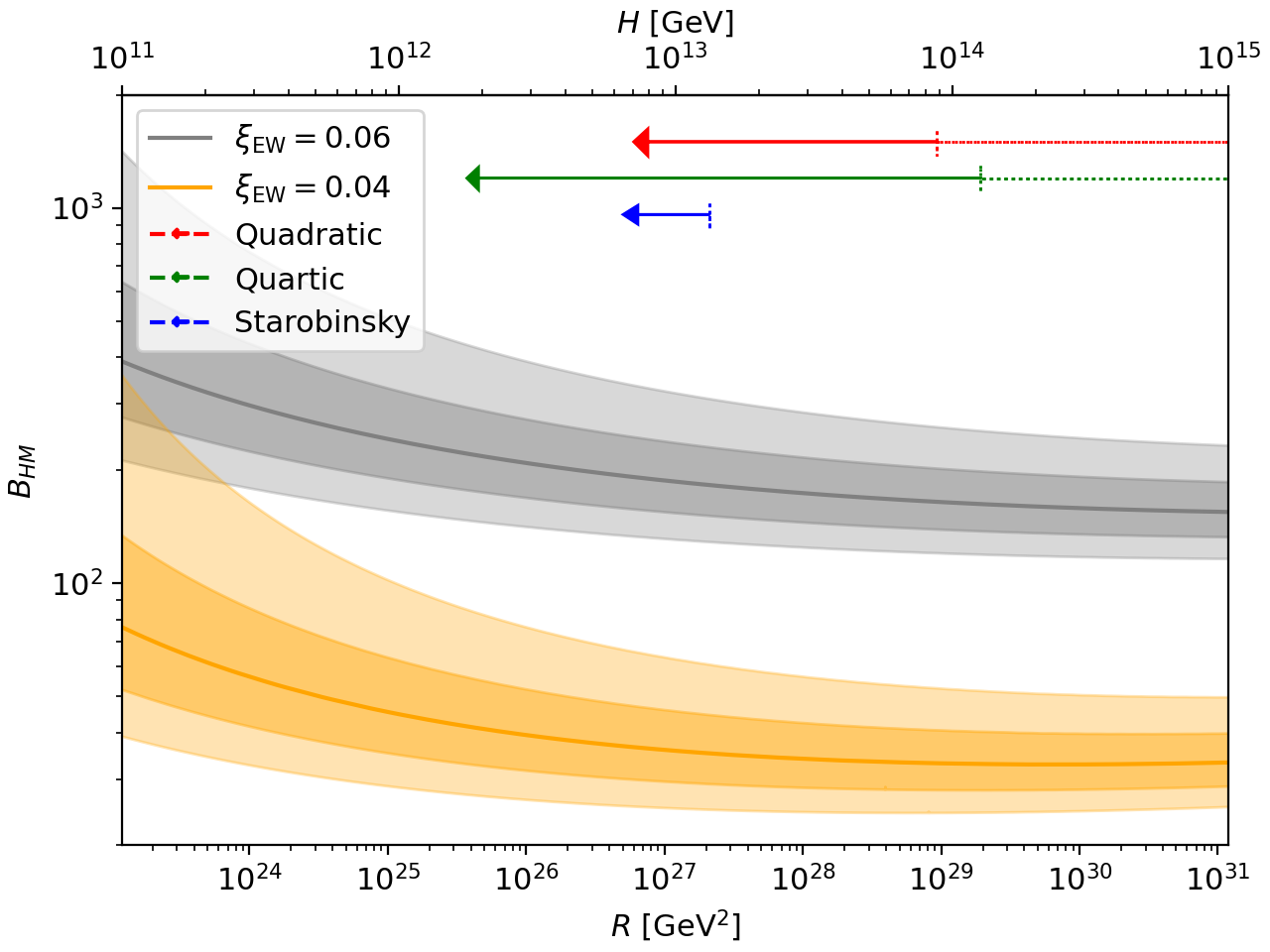}
    \caption{Hawking-Moss action difference $B_{\rm HM}$ for $\xi_{EW}=0.06, \, 0.04$ as a function of the Ricci scalar $R$ or the Hubble rate $(R=12H^2)$. The shaded areas denote $1\sigma$ and $2\sigma$ deviation from the central value of $m_t$, where a heavier top quark decreases the value of $B_{\rm HM}$ and vice versa. The solid red, blue and green arrows denote the last 60 e-foldings of inflation in quadratic, Starobinsky and quartic inflation respectively (see section \ref{subsec:inflationary-models}), whereas the dashed ones extend beyond that. }
    \label{fig:BHMs-H-R}
\end{figure}

In Fig.~\ref{fig:BHMs-H-R}, we see how the Hawking-Moss action $B_{\rm HM}$ defined in Eq.~(\ref{eq:BHM}) scales with the Ricci scalar/Hubble rate for different values of the curvature coupling and for a mass range of 2$\sigma$ for the top quark. The choice of the $\xi_{\rm EW}$ values shown is motivated by the bounds obtained in Section~\ref{sec:Results} as the ones to be of interest and they parallel the corresponding ones in Fig.~\ref{fig:ksis-mu}. The three arrows denote the last 60 $e$-foldings of each inflationary model, where their dashed tails extend to earlier times. Hence, the range of the $B_{\rm HM}$ values of interest for this study is the one coinciding with the arrow of the chosen model. We observe that as the top quark mass increases the action becomes less sensitive to the Ricci scalar $R$. At high $R$, the action becomes approximately constant, in line with the analytic approximation~(\ref{eq:BHM-approx}).

\section{Bubble nucleation during inflation} \label{sec:Bubble nucleation during inflation}

\subsection{Expected number of bubbles}

The observation that the Universe around us is still in the metastable phase implies that no bubble nucleation event took place in our past light-cone. For this to be compatible with the theory, it should predict that the probability ${\cal P}({\cal N})$ that there were ${\cal N}$ bubble nucleation events in our past light-cone must satisfy ${\cal P}(0)\sim 1$. When ${\cal N}$ is small, this probability should follow Poisson distribution, and therefore we can relate it to the expected number of bubbles $\langle {\cal N}\rangle$ through
\begin{equation}
    {\cal P}(0)=e^{-\langle {\cal N}\rangle}.
\end{equation}
Therefore, observations require $\langle{\cal N}\rangle \lesssim 1$.

The expected number of bubbles $\Nbub$ of true vacuum in our past light-cone is given by the integral~\cite{Markkanen:2018pdo} 
\begin{eqnarray}
\Nbub = \int_{\rm past} d^4x \sqrt{-g} \Gamma(x) \, .
\end{eqnarray}
The subscript ``past'' indicates that the integral is taken over the past light-cone. In a Friedmann-Robertson-Walker universe, 
\begin{equation}
ds^2=a(\eta)^2\left(d\eta^2-d\vec{x}^2\right),    
\end{equation}
where $\eta$ is the conformal time and $a(\eta)$ is the scale factor,
the comoving radius of the past light-cone at conformal time $\eta$ is $r(\eta)=\eta_0-\eta$, where $\eta_0$ is the conformal time today.
The expected number of bubbles nucleated between the start of inflation $\eta_{\rm start}$ and the end of inflation $\eta_{\rm inf}$ is therefore given by
\begin{eqnarray}
    \Nbub = \frac{4\pi}{3}\int_{\eta_{\rm start}}^{\eta_{\rm inf}} d\eta a(\eta)^4 (\eta_0-\eta)^3 \Gamma(a(\eta)) \,.
\end{eqnarray}

Instead of conformal time $\eta$, it is convenient to express this as an integral over the number of $e$-foldings $N=\ln (a_{\rm inf}/a(\eta))$, where  $a_{\rm inf}=a(\eta_{\rm inf})$ is the scale factor at the end of inflation.\footnote{Note that with this definition, higher $N$ corresponds to earlier time during inflation.} 
This gives the expected number of bubbles as a function of $N_{\rm start}$, i.e. the value of $N$ at the start of inflation or equivalently the total number of $e$-foldings of inflation,
\begin{eqnarray}
    { \Nbub(N_{\rm start}) =  \int_0^{N_{\mathrm{start}}} dN 
    \frac{4\pi}{3 H(N)}\left( \frac{a_{\mathrm{inf}} \left[\eta_0-\eta\left(N\right)\right]}{e^{N}} \right)^3 \Gamma(N) } \, ,
    \label{eq:Nbub-N}
\end{eqnarray}
where $H$ is the Hubble rate and it is given by
\begin{equation}
    H(N)=-\frac{1}{a_{\rm inf}}e^N\frac{dN}{d\eta} \,.
\end{equation}

For future reference, it is instructive to note that the integrand in Eq.~(\ref{eq:Nbub-N}), which we will denote by $\gamma(N)$, is a product of a geometric factor
\begin{eqnarray}
\frac{d{\cal V}}{dN} = \frac{4\pi}{3H(N)} \left(\frac{a_{\mathrm{inf}} \left[\eta_0-\eta\left(N\right)\right]}{e^{N}} \right)^3 \,  ,
\end{eqnarray}
and a dynamical factor $\Gamma (N)$, i.e.,
\begin{equation}
    \gamma(N) \equiv \frac{d\Nbub}{dN}=\frac{d{\cal V}}{dN}\Gamma(N).
    \label{eq:factors}
\end{equation}
The dynamical factor depends on the space-time geometry only through the Ricci scalar and is given by Eq.~(\ref{eq:Gamma}), whereas the geometric factor depends on the inflationary model.

\subsection{Numerical solution}

For a general single-field inflationary model, the space-time geometry is determined by the Friedmann and field equations
\begin{eqnarray}
H^2&=&\frac{1}{3M_P^2}\left(\frac{1}{2}\dot\phi^2+V(\phi)\right),\nonumber\\
\ddot\phi&=&-3H\dot\phi-V'(\phi),
\end{eqnarray}
where the dot indicates a derivative with respect to the physical time $t$, $\phi$ is the inflaton field and $V(\phi)$ is its potential. By solving these, one can obtain $\phi(N), H(N)$ and $\eta(N)$ and hence calculate the integral (\ref{eq:Nbub-N}) by using the bubble nucleation rate $\Gamma(N)$.

In practice, we rewrite these equations using $N$ as the time variable as a set of three coupled differential equations
\begin{eqnarray}
  \frac{d\Nbub}{dN} &=& \gamma(N)= \frac{4\pi}{3} \left[a_{\mathrm{inf}} \left(\frac{3.21e^{-N}}{a_0 H_0} -\Tilde{\eta}(N)\right)\right]^3 \frac{\Gamma(N)}{H(N)}  \, ,
  \label{eq:dNbubs-dN}\\
  \frac{d \Tilde{\eta}}{dN} &=& - \Tilde{\eta}(N) - \frac{1}{a_{\mathrm{inf}} H(N)} \,,
    \label{eq:detatilda-dN}\\
\frac{d^2\phi}{dN^2}
&=&\frac{V(\phi)^2}{M_P^2H^2}
\left(\frac{d\phi}{dN}-M_P^2\frac{V'(\phi)}{V(\phi)}\right) \, ,
\label{eq:dphi-dN}
\end{eqnarray}
where $\tilde{\eta}=e^{-N}\eta$ and the Hubble rate $H$ is given by
\begin{equation}
H^2 =  \frac{V(\phi)}{3M_P^2} \left[ 1 -  \frac{1}{6 M_P^2} \left( \frac{d \phi}{d N} \right)^2 \right]^{-1} \, .
\label{eq:Hubble-N}
\end{equation}
For the nucleation rate $\Gamma(N)$, given by Eq.~(\ref{eq:Gamma}), we need the Ricci scalar, given by
\begin{eqnarray}
    R = 12H^2 \left[ 1 -  \frac{1}{4 M_P^2} \left( \frac{d \phi}{d N} \right)^2 \right] \, .
\end{eqnarray}
Note that these equations do not assume the slow-roll conditions.

We solve this set of equations numerically using Mathematica, starting from slow-roll initial conditions at a high field value, which corresponds to a large value of $N$. To be precise, we fix the initial field value to that corresponding to $N=10^6 \, e$-foldings before the end of inflation, as calculated in the slow-roll approximation. We then integrate the equations down towards lower $N$, until the field reaches the point $\phi_{\rm inf}$ at which the expansion of the universe no longer accelerates, 
\begin{equation}
\left.\frac{\ddot{a}}{a}\right|_{\phi=\phi_{\rm inf}} =
\left.H^2\left[ 1 - \frac{1}{2 M_P^2} \left( \frac{d \phi}{d N} \right)^2 \right]\right|_{\phi=\phi_{\rm inf}}=0.
\end{equation}
We then set this point as the origin of our $N$ axis, i.e. $N(\phi_{\rm inf})=0$.

\subsection{Inflationary Models} \label{subsec:inflationary-models}

Within the slow-roll approximation, for a given model of inflation with the potential $V(\phi)$, the power spectrum of curvature perturbations has the expression~\cite{Liddle:2000cg}
\begin{equation}
{\cal P}_\zeta(k) = \f{V(\phi)}{24\pi^2 M_{\rm P}^4\epsilon}\,;\qquad \epsilon\equiv \f{M_{\rm P}^2}{2}\bigg(\f{V'(\phi)}{V(\phi)}\bigg)^2\,.\label{eq:CMB}
\end{equation}
Current CMB observations set the amplitude of the power spectrum to be ${\cal P}_\zeta(k_*)\approx2.1\times10^{-9}$ at the pivot scale $k_*=0.05{\rm Mpc}^{-1}$~\cite{Akrami:2018odb}, where the scale factor is chosen to be $a_0=1$ today. When precisely the scale corresponding to the pivot scale exits the horizon during inflation depends on the cosmic history and in particular on the reheating epoch. However, when determining the input parameters for inflationary models, we will assume that this takes place precisely 60 $e$-folds before the end of inflation, which in turn we take to occur when the potential slow-roll parameter satisfies $\epsilon=1$. All inflationary models studied in this work involve just one parameter and are then completely determined once the correct amplitude has been fixed.

In {\it quadratic inflation}, the inflaton potential has only a quadratic term 
\begin{eqnarray}
   V(\phi) = \frac{1}{2}m^2\phi^2 \, ,
   \label{eq:V-quad}
\end{eqnarray} where $m = 1.4 \times 10^{13}$ GeV acts as the inflaton mass and it is constrained from CMB measurements and Eq.~(\ref{eq:CMB}).

As the name suggests for {\it quartic inflation} one has
\begin{eqnarray}
    V(\phi) &=& \frac{1}{4} \lambda \phi^4 \, , 
    \label{eq:V-quar}
\end{eqnarray}
where $\lambda = 1.4 \times 10^{-13}$ comes from Eq.~(\ref{eq:CMB}).

The quadratic and quartic models do not provide a very good fit to data~\cite{Akrami:2018odb}, but we still consider them because of their simplicity. Starobinsky inflation \cite{Starobinsky:1979ty, Starobinsky:1980te} on the other hand complies with observational data very well and can draw connections between different inflationary models \cite{Liddle:2000cg, Kehagias:2013mya}. In this work, we consider a Starobinsky-like power-law model with the potential
\begin{eqnarray}
  V(\phi) &=& \frac{3}{4} \alpha^2 M_P^4 \left( 1 - e^{-\sqrt{\frac{2}{3}}\frac{\phi}{M_P}} \right)^2\,,
  \label{eq:V-star}
\end{eqnarray}
where the CMB fixes $\alpha = 1.1 \times 10^{-5}$ from Eq.~(\ref{eq:CMB}). This potential does not include cross-couplings between the Higgs and the inflaton, which in the Starobinsky model are introduced when the initial Lagrangian is parametrized in terms of an additional $R^2$-term \cite{Ema:2017rqn,He:2018gyf}, so it is not strictly speaking Starobinsky inflation. However, we will refer to it as {\it Starobinsky inflation} here for brevity.  

\section{Results} \label{sec:Results}

\subsection{Bounds on \texorpdfstring{$\xi$}{} }

Assuming a given inflationary model, and given values of the top quark mass $m_t$ and other Standard Model parameters, one can obtain a lower bound on the Higgs-curvature coupling $\xi$ by solving Eqs.~(\ref{eq:dNbubs-dN})--(\ref{eq:dphi-dN}) and requiring that $\langle {\cal N}\rangle \le 1$. The precise bound will depend on the duration of inflation, which we will discuss later in more detail. Because in Eq.~(\ref{eq:dNbubs-dN}), $\gamma(N)>0$, the longer inflation lasts, the stricter the bound on $\xi$. 

\begin{figure}[t]
    \centering
    \includegraphics[scale=0.7]{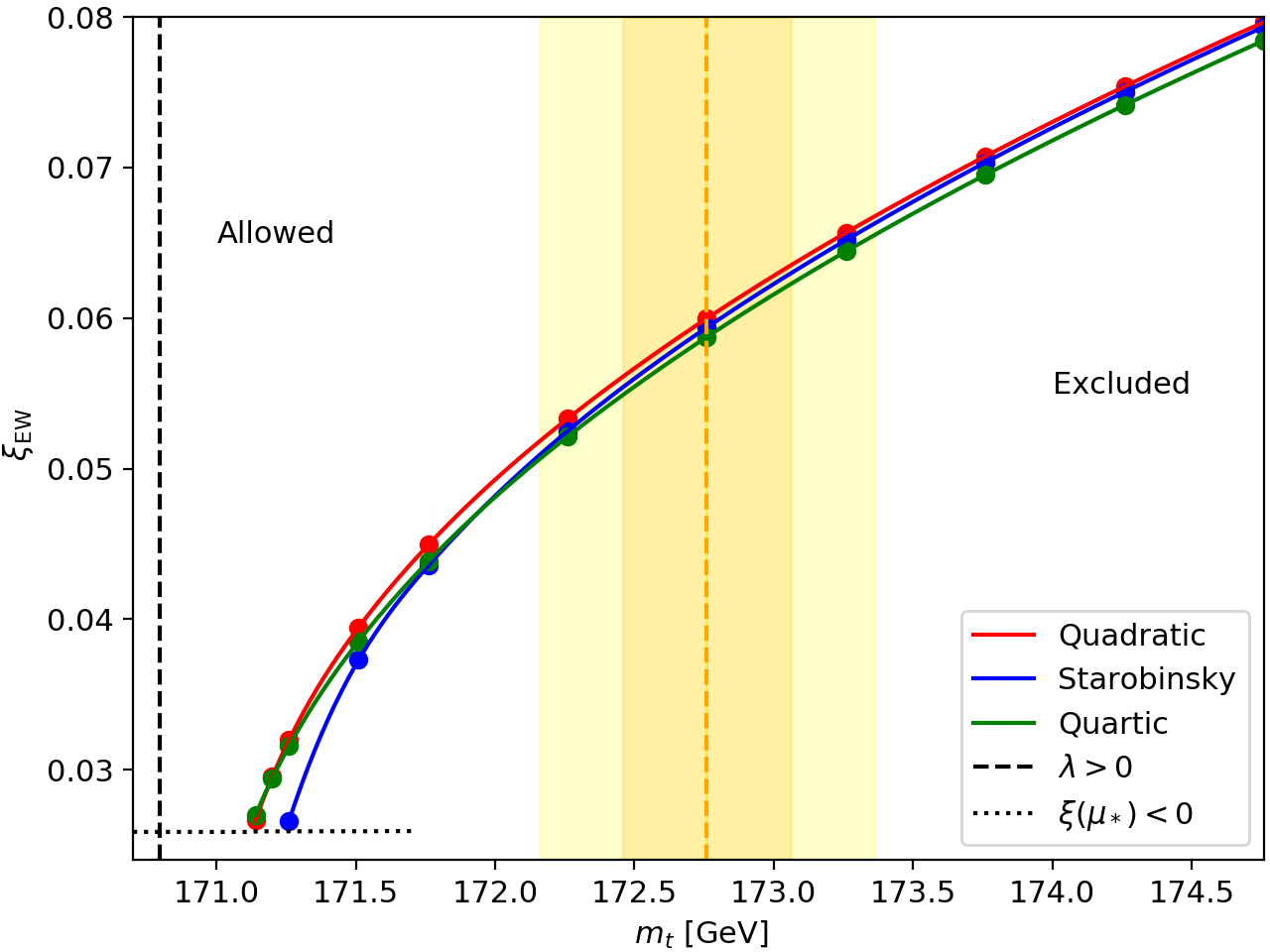}
    \caption{Constraints on the value of the curvature coupling $\xi$ at $\mu_{\rm EW}$ by imposing $\Nbub (60) \approx 1$ at the bound with a varying top quark mass for different inflationary models. The vertical dashed black line signifies the top quark mass threshold below which the self coupling of the Higgs field remains positive as it runs and thus prohibiting the formation of a second minimum in the Higgs potential. The vertical dashed orange line lies at the central value of $m_t = (172.76 \pm 0.30)$ GeV, where the shaded areas denote the corresponding $\pm \sigma$ and $\pm 2 \sigma$ variances \cite{Zyla:2020zbs}. The horizontal dotted black line shows the lowest $\xi_{\rm EW}$ value below which $\xi(\mu_*)$ turns negative as it runs.}
     \label{fig:ksi-mtop}
\end{figure}

In Fig.~\ref{fig:ksi-mtop} we show these lower bounds calculated for the three inflationary models discussed in Section~\ref{subsec:inflationary-models} and different values of $m_t$, based on the minimal assumption that inflation lasts $N=60 \,\, e$-foldings. We can see that all three inflationary models lead to very similar bounds, which indicates that, at least to some extent, they can be considered to be model-independent.

On the other hand, the bound depends quite significantly on the mass of the top quark. If it is sufficiently low, as indicated by the vertical black dashed line, the bound disappears completely because the electroweak vacuum becomes the true minimum. However, already at $m_t\approx 171.2~{\rm GeV}$, instability requires negative $\xi$ at the relevant scale $\mu_*$, as indicated by the horizontal dotted line. With a negative $\xi(\mu_*)$, the Higgs field gets displaced from its electroweak value during inflation. That changes its dynamics so much that we cannot use the same estimate (\ref{eq:dNbubs-dN}) for the expected number of bubbles, and more work is required to determine the actual constraints. Therefore we terminate the curves at that line. 

Finally, for completeness, we state explicitly the $\xi$-bounds for $m_t \pm 2 \sigma$ in each model
\begin{eqnarray}
\label{eq:xibounds0}
   \mathrm{Quadratic :} \,\,\, \xi_{\rm EW} \geq 0.060^{+0.007}_{-0.008}, \\
    \mathrm{Quartic :} \,\,\, \xi_{\rm EW} \geq 0.059^{+0.007}_{-0.008}, \\
    \mathrm{Starobinsky :} \,\,\, \xi_{\rm EW} \geq 0.059^{+0.007}_{-0.009},\label{eq:xibounds2}
\end{eqnarray}
where the numerical errors in the $\xi_{\rm EW}$'s for a fixed top quark mass are approximately $<1\%$ of their values (i.e. $\Delta \xi_{\rm EW} \approx 10^{-5} - 10^{-4}$).

\subsection{Bubble nucleation time}

In addition to the overall constraint on $\xi$, it is instructive to calculate the time during inflation at which bubbles are most likely to nucleate. This is important for two reasons: First, if bubbles were predominantly nucleated very close to the end of inflation, for example during the last $e$-folding, it would suggest that the constraints in Fig.~\ref{fig:ksi-mtop} may not be reliable. This is because  Eq.~(\ref{eq:Gamma})  is calculated in dS space-time, and near the end of inflation the space-time geometry deviates increasingly from dS. Second, if bubble formation was most likely to happen early on during inflation, before the last 60 $e$-foldings, then the bounds in Fig.~\ref{fig:ksi-mtop} would depend significantly on the early stages of inflation.

\begin{figure}[t]
    \centering
    \includegraphics[scale=0.65]{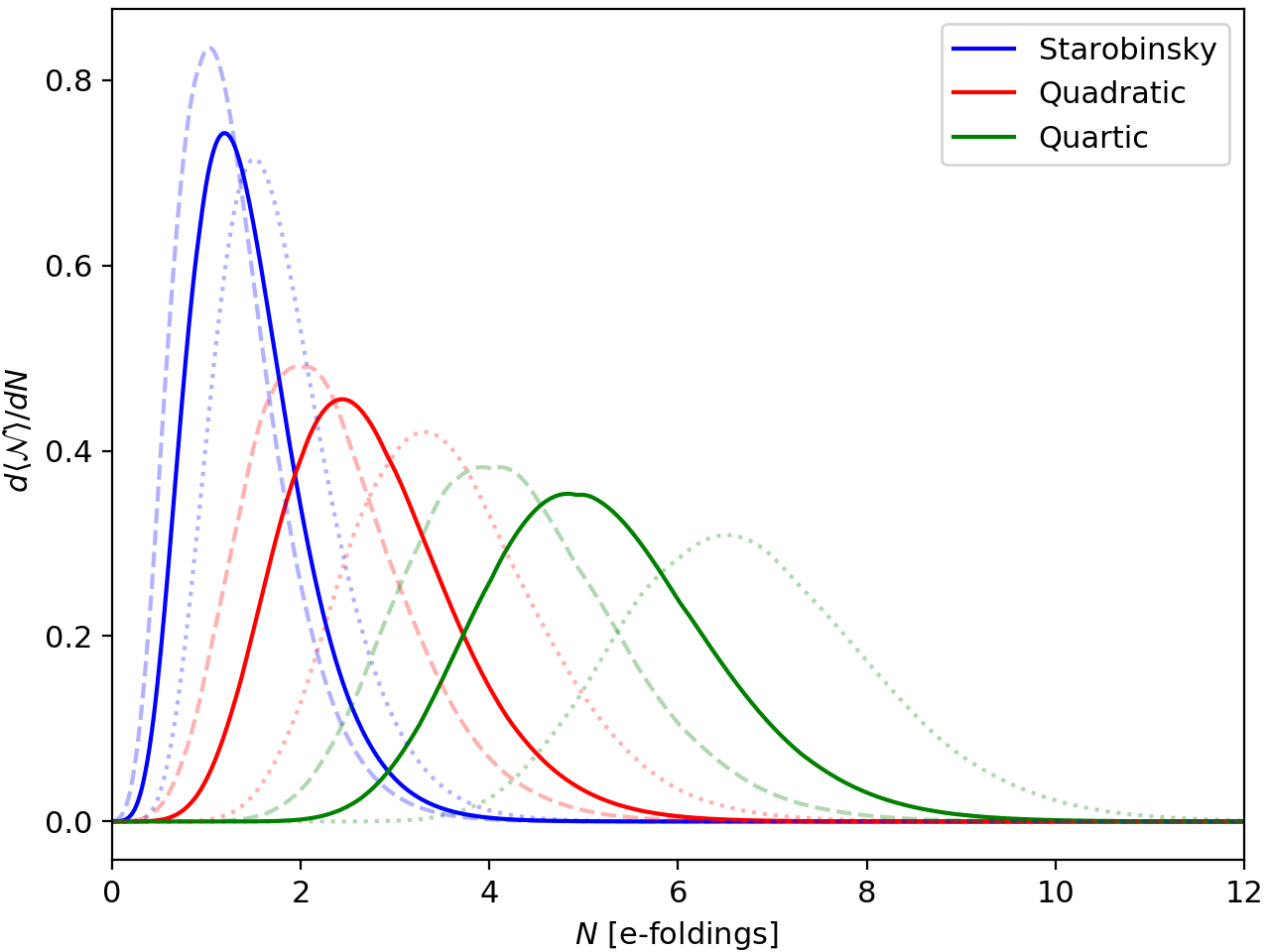}
    \caption{The integrands of the expectation value (\ref{eq:Nbub-N}) of the number of true vacuum bubbles $\Nbub$ as functions of $e$-foldings $N$ for the three different inflationary models, with $\xi_{\rm EW}$ chosen such that $\Nbub (60)=1$ in each case. This means that $\xi_{\rm EW}^{\rm Star} = 0.05938$, $\xi_{\rm EW}^{\rm Quad} = 0.05998$ and $\xi_{\rm EW}^{\rm Quar} = 0.05875$ for Starobinsky, quadratic, and quartic inflation, respectively.
    The solid lines correspond to the central value of $m_t$, whereas the dashed and dotted lines correspond to a deviation of $\pm 0.5$ GeV, respectively. The corresponding scales of the Hubble rates at which the bubbles are predominantly produced are $H_{\rm Star} = 9.96 \times 10^{12}$ GeV, $H_{\rm Quad} = 1.83  \times 10^{13}$ GeV and $H_{\rm Quar} = 1.16 \times 10^{13}$ GeV.}
    \label{fig:BubsPrime-N}
\end{figure}

In Fig.~\ref{fig:BubsPrime-N}, we show the probability $\gamma(N)$ of bubble nucleation per $e$-folding for the three inflationary models. In each case, $m_t$ has been assumed to have its experimental value, and $\xi_{\rm EW}$ is fixed to the value that gives $\langle{\cal N}\rangle=1$.
We can see that in all three cases, the function has a clear localised peak. This means that there is a definite, fairly well-defined time during inflation, when the vacuum decay is most likely to happen. In quadratic and quartic models, this peak is a few $e$-foldings before the end of inflation, which means that the constraints on $\xi$ should be reliable, and even in Starobinsky inflation it is more than one $e$-folding before the end. Also note that a lighter top quark ``pushes'' the peak to earlier times, while a heavier one towards the end of inflation.

\begin{figure}[b]
    \centering
    \includegraphics[scale=0.62]{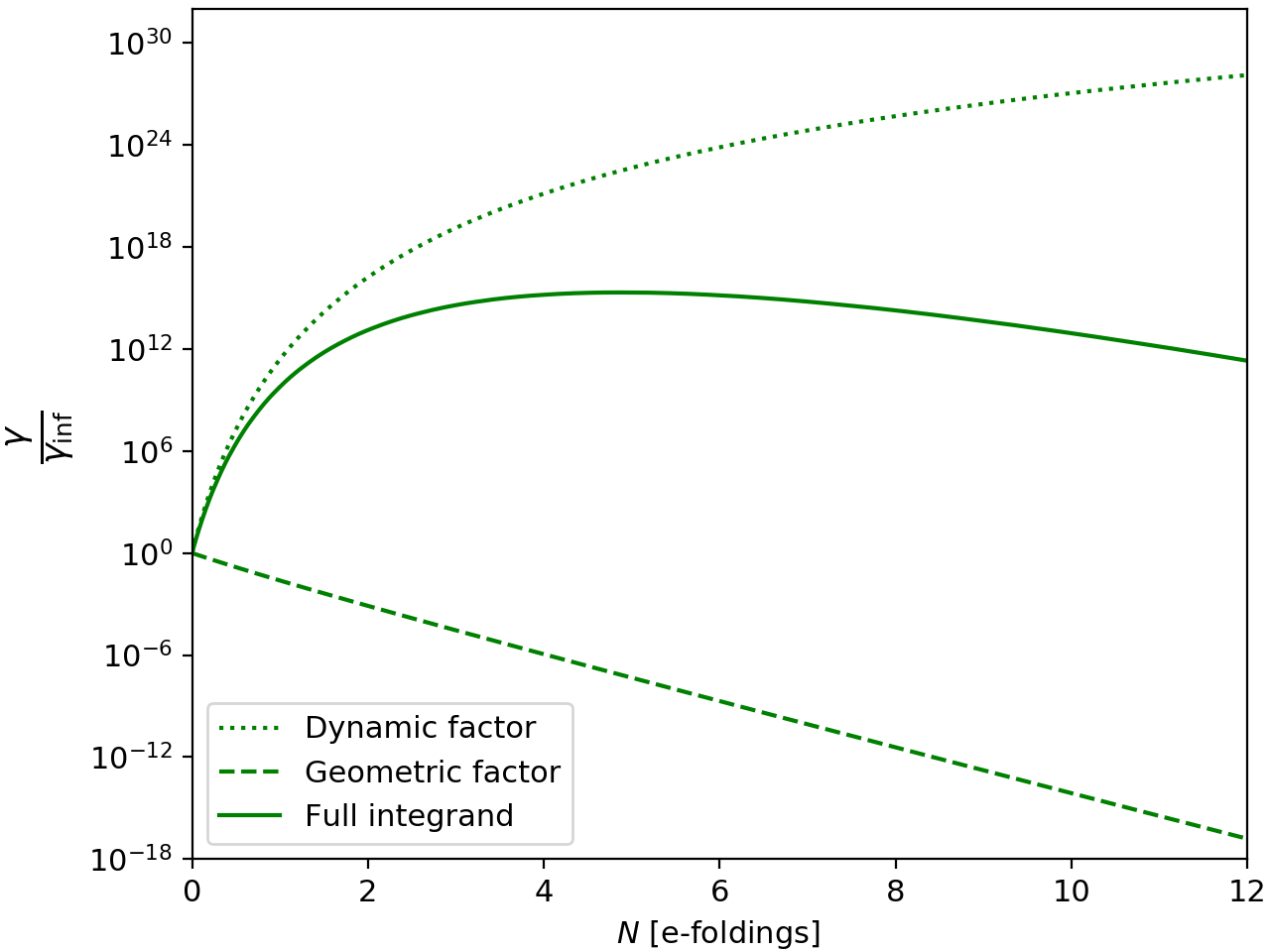}
    \caption{Factors of the integrand $\gamma(N)$ defined in Eq.(\ref{eq:factors}) as functions of $e$-foldings of inflation in quartic inflation with $\xi_{\rm EW} = 0.05875$ and $m_t=172.76$ GeV. The dynamic factor corresponds to $\Gamma(N)$, while the geometric factor to $\frac{d{\cal V}}{dN}$ and we have normalized all factors to one at $N=0$.}
     \label{fig:Integrand-factors-quartic}
\end{figure}

The reason for this localised peak can be see in Fig.~\ref{fig:Integrand-factors-quartic}. As pointed out in Eq.~(\ref{eq:factors}), the overall probability $\gamma(N)$ consists of two factors, the dynamical one and the geometric one. Because of the expansion of space, the geometric factor decreases exponentially as a function of $N$. The dynamical factor $\Gamma(N)$ increases, but not exponentially. Therefore the product $\gamma(N)$ has a maximum.

\subsection{Significance of the total duration of inflation} \label{subsec:duration}

As discussed in the previous section and is evident from Fig.~\ref{fig:BubsPrime-N}, bubble nucleation is strongly dominated by the dynamics close to the end of inflation. This implies that bounds from vacuum stability are relatively insensitive to the total duration of inflation as long as it is larger than around ten $e$-folds. However, in principle inflation can last for many orders of magnitude longer than this, and therefore it is important to check whether the behaviour at very large $N$ can change this conclusion.

To consider a long period of inflation, we split the integral (\ref{eq:Nbub-N}) into two pieces,
\begin{equation}
\langle{\cal N}\rangle(N_{\rm start}) = \langle{\cal N}\rangle(60) + \int_{60}^{N_{\rm start}} dN \, \gamma(N) ,
\label{eq:Nsplit}
\end{equation}
where we have already computed the first term numerically. If we choose the parameter values at the threshold, then by definition
$\langle{\cal N}\rangle(60)=1$. The second term, on the other hand, can be computed using the  slow roll approximation, which is valid at early times.

For quadratic inflation (\ref{eq:V-quad}), to leading order in slow-roll, we can solve
Eqs.~(\ref{eq:detatilda-dN})--(\ref{eq:Hubble-N}), analytically to obtain
\begin{eqnarray}
    \phi(N) &=& \left(M_P \sqrt{2}\right) \sqrt{1 + 2N}  \approx 2 M_P \sqrt{N} \, , \\
    H(N) &=& (m/\sqrt{3}) \sqrt{1 + 2N} \approx m \sqrt{\frac{2N}{3}}\, ,
    \label{eq:H-N-quad} \\
    \eta_0 - \eta(N) &=&  \frac{3.21}{a_0H_0} + \frac{1}{a_{\mathrm{inf}}m} \sqrt{\frac{3\pi}{2e}} \left[ \mathrm{erfi}\left( \sqrt{N + \frac{1}{2} }\right) - \mathrm{erfi}\left( \sqrt{\frac{1}{2}}\right) \right] \nonumber\\
&\approx& \sqrt{\frac{3}{2}}\frac{1}{ma_{\rm inf}}\frac{e^N}{\sqrt{N}}.
    \label{eq:eta-N-quad}
\end{eqnarray}
where the approximate forms are valid at $N\gg 1$.
The first term in Eq.~(\ref{eq:eta-N-quad}) is the post inflationary contribution \cite{Markkanen:2018pdo}, with $a_0$ and $H_0$ denoting the scale factor and the Hubble rate today, which is subdominant when $N\gtrsim 60$. Based on Eq.~(\ref{eq:BHM-approx}) and as implied also by Fig. {\ref{fig:BHMs-H-R}}, we assume that the Hawking-Moss action $B_{\rm HM}$ is approximately constant at early times, so that $\Gamma(N)\approx H(N)^4 e^{-B_{\rm HM}}$. Under these approximations, Eq.~(\ref{eq:Nsplit}) simplifies to
\begin{equation}
\langle{\cal N}\rangle(N_{\rm start})
\approx 1+\frac{4\pi e^{-B_{\rm HM}}}{3}N_{\rm start}.
\label{eq:asympt}
\end{equation}
This is true for all monomial potentials. Therefore, the early-time contribution is only important if $N_{\rm start}\gtrsim e^{B_{\rm HM}}\sim 10^{58-69}$ $e$-foldings depending on the value of $m_t \pm 0.5$ GeV and the corresponding $\xi_{\rm EW}$. For comparison, Eq.~(\ref{eq:H-N-quad}) shows that the requirement that the Hubble rate does not reach trans-Planckian values, $H(N_{\rm start})\ll M_P$, implies
\begin{equation}
    N_{\rm start}\ll \frac{3}{2}\left(\frac{M_P}{m}\right)^2\approx 10^{10}.
\end{equation}
Therefore, one can conclude that in quadratic inflation, early bubble production is never important.

The same conclusion holds in the quartic model (\ref{eq:V-quar}).
The slow-roll solution for the inflaton, Hubble rate and conformal time is
\begin{eqnarray}
  \phi(N) &=& \left(2 M_P \sqrt{2}\right) \sqrt{1 + N} \approx 2 M_P \sqrt{2N} \, , \\
   H(N) &=& 4 M_P \sqrt{\frac{\lambda}{3}} (1 + N) \approx \left( 4 M_P \sqrt{\frac{\lambda}{3}} \right) N , \\
   \eta_0 - \eta(N) &=& \frac{3.21}{a_0H_0} + \frac{\sqrt{3/\lambda}}{4 M_P a_{\mathrm{inf}} e} \left[ {\rm Ei}(N+1) - {\rm Ei}(1) \right] \approx \left( \frac{\sqrt{3/\lambda}}{4 M_P a_{\mathrm{inf}}} \right) \frac{e^N}{N},
\end{eqnarray}
where ${\rm Ei}$ is the exponential integral function. The asymptotic result for the number of bubbles is again Eq.~(\ref{eq:asympt}), and the constraint arising from trans-Planckian Hubble rates is
\begin{equation}
    N_{\rm start}\ll\frac{1}{4} \sqrt{\frac{3}{\lambda}} \approx 10^6.
\end{equation}

For the plateau models such as the Starobinsky-type potential in Eq.~(\ref{eq:V-star}) the energy density quickly approaches a constant for large $N_{\rm start}$ and thus remains strictly sub-planckian. Therefore it does not give rise to an upper limit on the total number of $e$-foldings. The solution of Eqs.~(\ref{eq:detatilda-dN})--(\ref{eq:Hubble-N}) is
\begin{eqnarray}
   \phi(N) &=&  M_P \sqrt{\frac{3}{2}} \left[ \mathrm{ln}f -f -\frac{4}{3}N - W_{-1} \left[ - e^{\mathrm{ln}f -f -\frac{4N}{3}} \right] \right]  \approx M_P \sqrt{\frac{3}{2}} \mathrm{ln}N , \\ 
   H(N) &=& \frac{\alpha M_P}{2} \left[ 1 - \frac{1}{f} e^{\left( f\sqrt{\frac{2}{3}} + \frac{4N}{3} + W_{-1} \left[ - e^{\mathrm{ln}f - f -\frac{4N}{3}} \right] \right)}   \right] \approx \frac{\alpha M_P}{2},\\
   \eta_0 - \eta(N) &\approx& \frac{3.21}{a_0H_0} + \frac{2\left(e^N -1\right)}{a_{\rm inf} \alpha M_P} \approx 
   \frac{2 e^N }{a_{\rm inf} \alpha M_P},
\end{eqnarray}
where $f=1+2/\sqrt{3}$, and $W_{-1}$ is the -1 branch of the Lambert function \cite{Martin:2013tda}. The asymptotic behaviour of the expected number of bubbles is, again, Eq.~(\ref{eq:asympt}),
and therefore if $ N_{\rm start}\gtrsim e^{B_{\rm HM}}\approx 10^{60}$, the vacuum stability bounds on $\xi$ become stronger. The same conclusion applies to all cases where the Hubble rate approaches a constant at high $N$.

\section{Conclusions} \label{sec:Conclusions}
 In this work we have obtained bounds on the Higgs-curvature coupling coming from vacuum instability in the SM during inflation. We considered three models of inflation, quadratic and quartic chaotic inflation and Starobinsky-like power-law inflation. For the calculation of the tunneling probability we made use of the Hawking-Moss bounce with the renormalization group improved effective potential calculated on a curved background without the usual assumption of strict de Sitter space.
 
 In the effective potential for the SM Higgs we included the leading time-dependent curvature corrections from all SM constituents by making use of the results of Ref.~\cite{Markkanen:2018bfx} and choosing the renormalization scale such that the loop correction strictly vanishes, as written in Eq.~(\ref{eq:dcond}). This amounts to a consistent inclusion of quantum induced curvature corrections to 1-loop order, which in previous works has not been addressed beyond de Sitter space making the bounds presented here arguably the most accurate to date. 
 
 Our analysis indicates that the bound for the Higgs non-minimal coupling at the electroweak scale is
 \begin{equation}
 \xi_{\rm EW} \gtrsim 0.06\,,
 \end{equation}
 for the central value of the top quark mass, valid for quadratic, quartic and Starobinsky-like inflationary models. This is numerically close to bounds obtained earlier in the de Sitter approximation \cite{Markkanen:2018pdo,Markkanen:2018bfx}. From the results in Eqs.~(\ref{eq:xibounds0})--(\ref{eq:xibounds2}) one can also see that the bound is largely unchanged when varying the top quark mass by $2\sigma$ or less. Since for the non-minimal coupling in the SM, $\xi=0$ is not a fixed point of the RG evolution, a small yet non-zero value is perfectly natural from the model building point of view.
 
 As mentioned in the introduction, the issue of vacuum stability during inflation was similarly studied on a time-dependent background in Ref.~\cite{Fumagalli:2019ohr}. The main difference to our work was making use of the stochastic approach of Ref.~\cite{Starobinsky:1994bd} instead of the Hawking-Moss bounce and including Planck-suppressed derivative operators in the action. Furthermore, Ref.~\cite{Fumagalli:2019ohr} made use of the RG improved tree-level potential with the scale $\mu^2=h^2+12H^2$ in contrast to our using the 1-loop result in Eq.~(\ref{eq:RGIPot}) with the scale choice in Eq.~(\ref{eq:dcond}). For the cases where the analyses overlap there is good agreement between the results.
 
In this work, we have not considered a direct coupling between the Higgs and the inflaton field. In the slow-roll limit, there are examples in which its effects are similar to those of the curvature coupling and therefore one can translate the bounds on the curvature coupling to include the direct Higgs-inflaton coupling~\cite{Ema:2017loe}. Unfortunately this is not possible in general beyond slow roll, and therefore a new calculation is required to include the effects of a direct coupling. We leave this for future work.
  
Our work also revealed non-trivial insights concerning when precisely during inflation the vacuum bubbles are formed: consistently in all the three models that we studied the bulk of the bubble nucleation occurs close to the end of inflation. As Fig.~\ref{fig:BubsPrime-N} shows the probability for nucleation peaks localised less than ten $e$-folds before the end of inflation, while dropping rapidly for large $N$. In Section~\ref{subsec:duration} the dependence of the average number of bubbles on the total length of inflation was studied analytically. It was shown that the results are largely insensitive to the entire duration of inflation unless one considers an extremely long period of primordial inflation lasting more than $10^{50} \, e$-folds. Although a very large number, this might be significant for cases admitting eternal inflation~\cite{Jain:2019wxo}, which warrants further study.

\section*{Acknowledgments}
We would like to thank Fedor Bezrukov and José Eliel Camargo-Molina for useful discussions and help with the numerical computations. 
AR was supported by STFC grants ST/P000762/1 and ST/T000791/1, and by an IPPP Associateship. AM was supported by an STFC PhD studentship.
This project has received funding from the European Union’s Horizon 2020 research and innovation programme under the Marie Sk\l odowska-Curie grant agreement No. 786564.

\bibliography{references.bib}
\end{document}